%% file: example.tex
\documentclass[10pt]{article}

\input{preamble.tex}

\usepackage{lmodern}

\usepackage{graphicx}
\usepackage{times}
\usepackage{adjustbox}
\usepackage{float}
\usepackage{gensymb}

\title{3DSiameseNet to Analyze Brain MRI}

\authorname{Cecilia Ostertag\textsuperscript{1-2}, Marie
  Beurton-Aimar\textsuperscript{2}, Thierry Urruty\textsuperscript{3}}
\authoraddr{for the Alzheimer's Disease Neuroimaging Initiative* \\ \textsuperscript{1} L3i - EA 2118 LaRochelle University,
  France, cecilia.ostertag1@univ-lr.fr  \\ \textsuperscript{2} LaBRI -
  CNRS 5800 Bordeaux University, France, beurton@labri.fr
  \\ \textsuperscript{3} Univ. Poitiers, CNRS, XLIM, UMR 7252, F-86000
  Poitiers, France, thierry.urruty@univ-poitiers.fr}

\begin{document}
\noindent

\maketitle

*{\small{Data   used   in   preparation   of   this   article   were
    obtained   from   the  Alzheimer’s   Disease
    Neuroimaging Initiative (ADNI) database (adni.loni.usc.edu). As
    such, the investigators within the ADNI contributed to the design
    and implementation of ADNI and/or provided data but did not
    participate in analysis or writing of this report. A complete
    listing of ADNI investigators can be found at: http://adni.loni.usc.edu/wp-
content/uploads/how\_to\_apply/ADNI\_Acknowledgement\_List.pdf}}

\keywords
\textit{deep learning, siamese network, brain MRI, neurodegenerative disease, convolutional neural network}

\abstract
Prediction of the cognitive evolution of a person susceptible to
develop a neurodegenerative disorder is crucial to provide an
appropriate treatment as soon as possible. In this paper we propose a
3D siamese network designed to extract features from whole-brain 3D
MRI images. We show that it is possible to extract meaningful features
using convolution layers, reducing the need of classical image
processing operations such as segmentation or pre-computing features
such as cortical thickness. To lead this study we used the
Alzheimer's Disease Neuroimaging Initiative (ADNI), a public data base
of 3D MRI brain images. A set of 247 subjects has been extracted, all
of the subjects having 2 images in a range of 12 months. In order to
measure the evolution of the patients states we have compared these 2
images. Our work has been inspired at the beginning by an article of  Bhagwat et al. in 2018, who
have proposed a siamese network to predict the status of patients but
without any convolutional layers and reducing the MRI images to a
vector of features extracted from predefined ROIs. We show that our network achieves
an accuracy of 90\% in the classification of cognitively declining VS stable
patients. This result has been obtained without the help of a
cognitive score and with a small
number of patients comparing to the current datasets size claimed in deep
learning domain.

\section{Introduction}
Neurodegenerative diseases like Alzheimer, Parkinson, \ldots are disorders resulting in brain
tissue damage. Because of the heavy impact on the brain morphology in
this kind of diseases, medical imaging techniques such as structural
MRI can be used to study them, and to distinguish between healthy
patients, patients affected by disease or patients with mild
cognitive impairment. This way, a computational  model fitted to this
task can use an MRI image to give a diagnosis. Beyond the diagnosis,
for a better care of affected people, it becomes necessary to predict
the evolution of the disease, to adapt the treatment of already
identified patients, and to identify 
which patients are at high risk and have to be watched
closely\cite{frisoni2010clinical}. 

To predict evolution of such
diseases in subjects with regular yearly medical examinations, we have
designed a 3D siamese neural network taking into account the MRI
images of two time 
points to predict their cognitive evolution. Our model is
inspired by  the work of Bhagwat et. al. \cite{bhagwat2018modeling} who
have proposed a siamese network to predict the evolution of patients
by using measures computed from MRI images and a cognitive score. On one hand, as
we would like to take into account 3D MRI, our network is made of
several layers of convolutional layers for each of its branches.  
. On the other hand, our task has been to find if there are
differences between two images of the same patients (at two
different times), so the layer merging the two siamese branches has been defined to extract
differences and not similarities as is usually the case with siamese models. In the
next sections, we will first detail the architecture of our network,
then we will present our results, and finally we will give insights to
the discriminative abilities of our model.

\section{Related works}

The success of Deep Neural Networks (DNNs) for image classification
tasks was shown by LeCun in 1998 \cite{lecun1998gradient}. DNNs are
multi-layer architectures designed to extract meaningful features in
data, in order to solve a given task. These models are trained and
optimized using gradient backpropagation: a loss function is computed
to quantify the error of the model at each iteration, and the training
phase aims to minimize this error by adjusting the model's parameters
in order to enhance its performance step by step. During training, the loss function is computed on a training dataset and on an independent validation dataset. The difference is that the network's weights are not updated when using validation data, so the validation loss evaluates the performance of the network on unknown data. This method is used to ensure that the model is not overfitting the data.

To process image data, variations of DNNs, called Convolutional Neural
Networks (CNNs) are most often used. They consist of a sequence of
convolution layers, followed by pooling layers. The successive
convolution operations select the most important features of the
images, thus replacing the computation of hand-crafted features, and
the pooling operations reduce the dimensions of the data. After the
convolution layers, a classifier made of fully connected (or dense)
layers, use these informations to predict classes.

To face the challenge of image pair comparison, another type of DNNs
are convolutive siamese networks \cite{koch2015siamese}. They are made
of two identical branches, that share the same weights at
initialization and during all training. A distance or similarity
function is then used to compare two inputs, and based on this the
network will be trained to distinguish similar pairs from dissimilar
pairs. Such networks have been used for example for face recognition
\cite{zagoruyko2015learning, lin2016homemade}.

MRI data are medical images representing in 3D the anatomy of an organ
or tissue. The standard DNN architectures used with this type of
volumetric data are CNNs. In the case of MRI brain images, for example
Alzheimer's Disease data, several input formats have been used to
train these networks: 2D slices from the MRI volumes
\cite{sarraf2016classification,billones2016demnet}, whole 3D volumes
\cite{hosseini2016alzheimer,payan2015predicting,korolev2017residual,khvostikov20183d},
or regions of interest (ROIs) selected using an atlas
\cite{khvostikov20183d}. Also, a 2D+ approach used the three
anatomical plane projections was proposed
\cite{aderghal2017classification}. Using 3D ROIs is effective to
reduce memory requirements, but implies \textit{a priori} knowledge of
the brain areas involved in the disease. The model that we present in
this work does not rely on ROI selection.

In all previously cited articles, the classification between the
different groups of patients is inferred from only one time point (one
MRI image), so the diagnostic that is predicted is only valid for that
time point. To predict the symptomatic evolution of Alzheimer's
Disease in subjects, Bhagwat et. al. \cite{bhagwat2018modeling}
present a deep learning architecture which uses MRI and clinical data
from two medical visits. To handle the two MRI images, they use a
siamese network and create a "distance embedding" to model the changes
between the two visits. The main drawback of this model is that it is
dedicated to use 1D vectors of measures computed from MRI data instead
of directly using MRI images. According to the work of Payan and
Montana \cite{payan2015predicting} a 3D approach has better
performances than a 2D approach because it is
able to extract 3D patterns. The model that we propose is a siamese
network as suggested by Bhagwat et. al. but with some major
differences to take into account input pairs of 3D images.

Our architecture is based on the siamese network part of Bhagwat
et. al. Logintudinal Siamese Network (LSN)
\cite{bhagwat2018modeling}. In this article the goal of the authors is
to predict the evolution of Alzheimer's disease using two classes,
"Stable" and "Decline". Obviously, "Stable" represents the subject
with no cognitive decline, and "Decline" the subjects with cognitive
decline. To create these classes, the authors used a clustering
technique to separate the data based on a cognitive score. The LSN
network is composed of two parallel branches, made of fully connected
layers, and take as input a pair of 1D vectors. The first vector
contains 78 cortical thickness values computed from the baseline visit
MRI of a subject, and the second vector contains 78 cortical thickness
values computed from the 12-month follow-up visit of the same
subject. To obtain these values, the MRI images are segmented in 78
ROIs, and the cortical thickness value of each ROI is estimated. At
the end of both branches, the two outputs are concatenated into
one. Contrarily to usual siamese networks, this model is not used to
evaluate the similarity between the two inputs to determine if they
are from the same person, but to evaluate the difference between the
two inputs to quantify the changes in the brain morphology induced by
the disease. As stated previously we want to use directly 3D
MRI images, so we have adapted the LSN model to process 3D inputs.

\section{Network model}

\subsection{Architecture}

Image processing operations as segmentation and computation of
cortical thickness steps are time consuming, often require arbitrary
selection of ROIs, and can be subject to errors. To avoid this, we
decided to adapt the LSN architecture \cite{bhagwat2018modeling} to whole MRI 3D inputs. As we faced
problems of memory and computation time due to the use of large and
tridimentional data, we had several choices to make during the network
implementation. Moreover, the choice of layers and parameter values
were also conditioned by the need to extract small local features that
are characteristic of changes in the brain.

\begin{figure}[ht]
\centering
\includegraphics[scale=0.20]{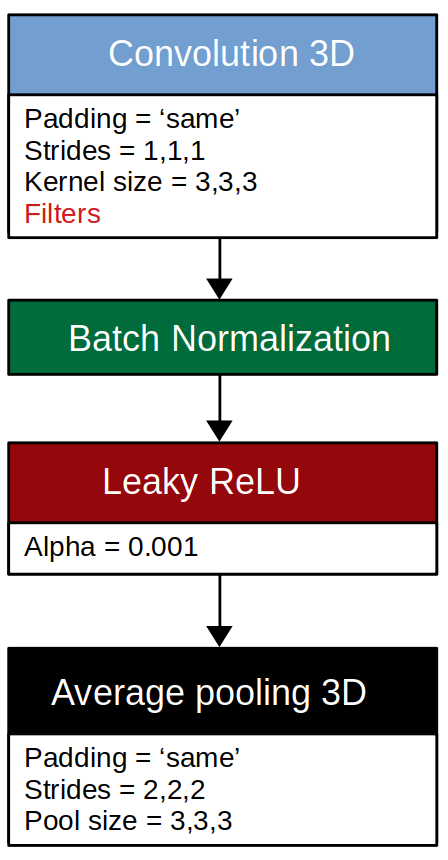}
\caption{\label{block}{\small{Elementary block: 1 Convolution 3D layer, with a
  variable number of filters, 1 Batch Normalization layer, 1 Leaky
  ReLU layer, 1 Average Pooling 3D layer.}}}

\end{figure}

The LSN model uses 1D inputs, so their siamese network is only made of
fully-connected layers. To use 3D data, we replaced the
fully-connected layers by 3D convolution layers, followed by 3D
pooling layers. For all convolution layers we chose a kernel size of
3x3x3 and a stride of 1 to preserve the details of the images. We used
average pooling instead of max pooling to avoid a bias of the network
towards edge detection. In Fig.\ref{block} we detail the elementary
block of our network.

\begin{figure}[H]
\centering
\includegraphics[width=9cm,height=\textheight]{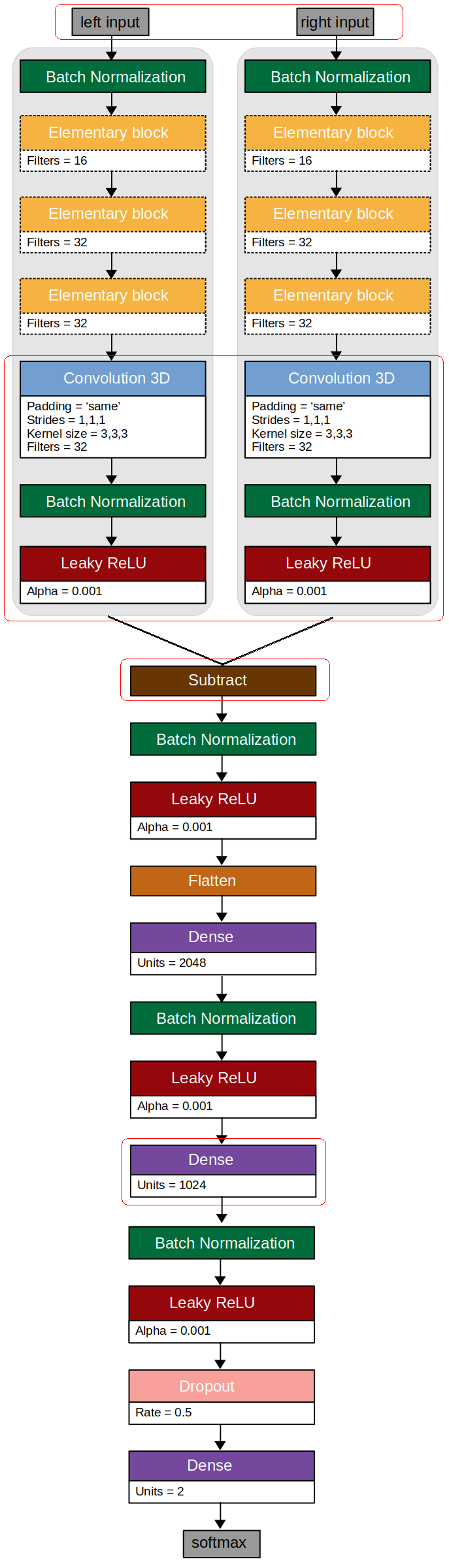}
\caption{\label{graph}{\small{3D-SiameseNet architecture. left input is the baseline MRI, and right input is the follow-up (12-month visit) MRI. Layers circled in red are studied in part 5.}}}
\end{figure}

It is made of one convolution layer, one batch
normalization layer \cite{ioffe2015batch}, one leaky ReLU
\cite{maas2013rectifier}, and one average pooling layer. This block is
repeated three times in each branch of our network.

Because of memory allocation problems, we used a small number of
filters for the convolution layers. The fusion of the two branches of
the network was done by computing the difference instead of using
concatenation, to have a better representation of the evolution of the
subject's brain morphology. Finally we used three fully-connected
layers, with softmax activation for the last layer. The complete
architecture of our 3D-SiameseNet is presented in Fig.\ref{graph}. To
avoid overfitting a dropout of 50 \% is applied before the output
layer. Each layer is followed by batch normalization and leaky ReLU to
improve the convergence of our model, as suggested by Lin
et. al. \cite{lin2016homemade}.

\subsection{Preprocessing and data augmentation}

\begin{figure}[H]
\centering
\includegraphics[width=8.5cm,height=3.5cm]{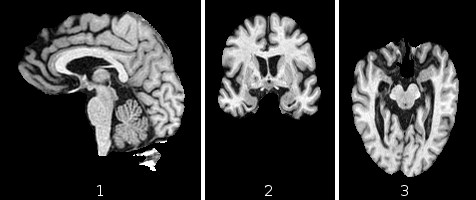}
\caption{\label{inputs}{\small{Example of brain image after skull-stripping:
  horizontal, frontal, and longitudinal slices from a subject's
  baseline MRI.}}}

\end{figure}

The data we used come from Alzheimer's Disease Neuroimaging Initiative
(ADNI) database \cite{noauthor_adni_nodate}. We used skull-stripping
\cite{skull} to remove features non related to brain morphology, and
zero-padding to ensure that all images have the same dimensions. The
pixel values were scaled between 0 and 1, and the images were
downscaled by 2 using an order 3 spline interpolation to reduce memory
allocation. The final dimensions of our inputs are 102x108x75. For
training we used pairs of images: the baseline MRI and the 12-month
follow-up MRI. (see Fig.\ref{inputs})

Our dataset contains 247 image pairs. We take 40 images for validation
and the rest in training. Because of the modest size of our training
dataset compared to deep learning requirements, we used a data
augmentation strategy. It consisted of random left/right rotation
between 0\degree and 5\degree, and left/right axial symmetry. During training,
data augmentation is applied image per image, at every epoch, on the
training images only.

\subsection{Training protocol}

To assign labels to our data, we followed the script provided by
Bhagwat et. al. \cite{bhagwat2018modeling}, subjects from the ADNI1
cohort were used for trajectory modelling, dividing the dataset
between "Stable" and "Decline" subjects, based on the Mini Mental State Evaluation clinical scores. We selected
a subset of these subjects in our dataset: 134 "Stable" subjects,
and 113 "Decline" subjects. For all of these subjects we have a
baseline and a 12-month MRI.

The model was trained with categorical crossentropy loss, optimized
with Adam \cite{kingma2014adam}, with a base learning rate of
0.001. To reduce overfitting we applied L2 regularization to all
kernel weights. We used 10 times random sub-sampling validation to
assess the bias induced by the choice of validation data. The model
was trained for 800 epochs, with the training data shuffled after
every epoch, and using a batch size of 20 due to memory
limitations. To evaluate the model, we used accuracy and Mean Squared
Logarithmic Error (MSLE).

The model was implemented using Keras \cite{chollet2015keras} with TensorFlow \cite{tensorflow2015-whitepaper} backend, and run on two NVIDIA 2080 Ti GPUs.
The source code can be found at
https://github.com/morphoboid/3D-SiameseNet

\section{Results}
\begin{figure}[h]
\centering
\includegraphics[width=9cm,height=5cm]{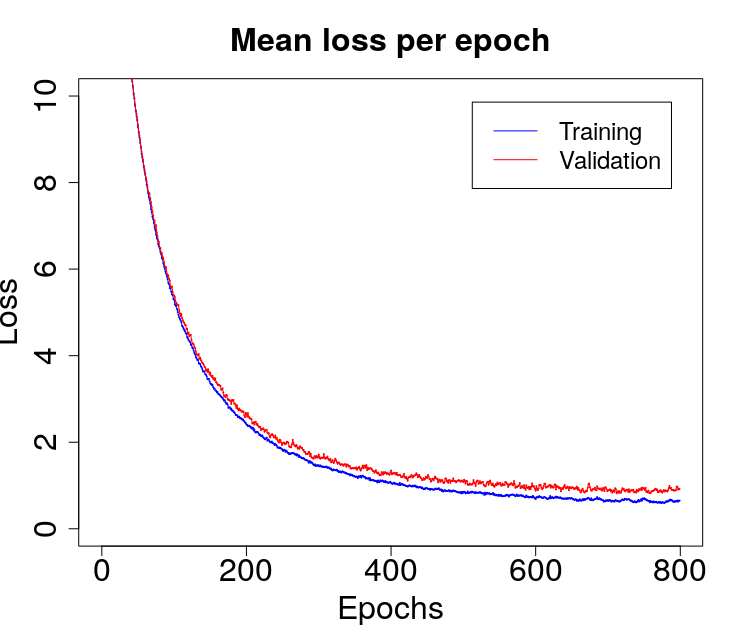}
\caption{\label{loss}{\small{ Evolution of the loss function for the training
  and validation sets over 800 epochs. Each point represents the mean
  of the loss values for ten random sub-sampling runs of the model, at
  the given epoch.}}}

\end{figure}

Our network's loss reaches a plateau after about 600 epochs (see
Fig.\ref{loss}), with a training loss of 0.624 and a validation loss
of 0.992 at the end of training.

We used accuracy to evaluate our model's ability to predict the right
class, and MSLE to evaluate the probabilities associated to these
predictions. After training, our model achieves a mean accuracy of 90\% and a mean MSLE of 0.04 (See Table \ref{tab2}).

\begin{table}[h]
\centering
\begin{adjustbox}{width=8.5cm,height=0.7cm}
\begin{tabular}{|p{1.5cm}|p{1.1cm}|p{1.5cm}|p{1.1cm}|p{1.2cm}|}
\hline
                    & \textbf{Mean Acc} & \textbf{Mean MSLE} & \textbf{Std Acc} & \textbf{Std MSLE} \\ \hline
\textbf{Training}   & 97\%              & 0.01      & 1.9\%           & 0.007     \\ \hline
\textbf{Validation} & 90\%               & 0.04      & 4.6\%            & 0.019      \\ \hline                                  
\end{tabular}
\end{adjustbox}
\caption{\small{{Mean and standard deviation for accuracy (Acc) and mean squared logarithmic error (MSLE) over 10 random sub-sampling validation runs of our model, after the last epoch.}}}
\label{tab2}
\end{table}

Bhagwat et. al. did performance evaluation experiments using 1116
subjects, to compare their results with four other models: a logistic
regression (LR), a support vector machine (SVM), a random forest (RF),
and an artificial neural network (ANN). These four models only use
imaging data (baseline MRI + follow-up MRI). Their LSM uses imaging
and clinical data from two time points, and achieved 94\% accuracy
(see Table \ref{tab3}). While we cannot at this time compare our
results to the LSM network because we only use imaging data, we show
that our model outperforms LR, SVM, RF, and ANN models, using
significantly less images.

\begin{table}[h]
\centering
\begin{adjustbox}{width=8.5cm,height=0.8cm}
\begin{tabular}{|p{1.5cm}|p{0.5cm}|p{0.7cm}|p{0.5cm}|p{0.7cm}|p{1.6cm}|p{2cm}|}
\hline
                    & \multicolumn{5}{c|}{\textbf{Imaging data only}}   & \textbf{Imaging and clinical data} \\ \hline 
                    & \textbf{LR} & \textbf{SVM} & \textbf{RF} & \textbf{ANN} & \textbf{3D-SiameseNet} & \textbf{LSM} \\ \hline
\textbf{Accuracy}   & 77\%            & 76\%      & 76\%           & 75\%   & 90\%   & 94\%   \\ \hline
\end{tabular}
\end{adjustbox}
\caption{{\small{Comparison of accuracy values given by six models: Logistic
  Regression (LR), Support Vector Machine (SVM), Random Forest (RF),
  Artificial Neural Network (ANN), Bhagwat et. al.'s Longitudinal
  Siamese Network (LSM), and our 3D-SiameseNet. Data from Bhagwat
  et. al. S4 File (Supplementary material). \cite{bhagwat2018modeling}}}
}

\label{tab3}
\end{table}

Along with a high accuracy, the MSLE over the validation set is low,
showing that the probabilities associated to the predictions are close
to the true labels. This should ensure a small number of false
positives and false negatives during inference.

As we mentioned in Section 3, we have been mainly limited by the
memory size of our GPU clusters. With more memory-expansive computing
resources, we could use MRI inputs at their original dimensions and
avoid downscaling artifacts. We could also add filters to the
convolution layers to extract more features. With this configuration,
we assume that our model's accuracy would further increase.

\section{Discriminative abilities of our model}

\begin{figure}[h]
\centering
\includegraphics[width=8.8cm]{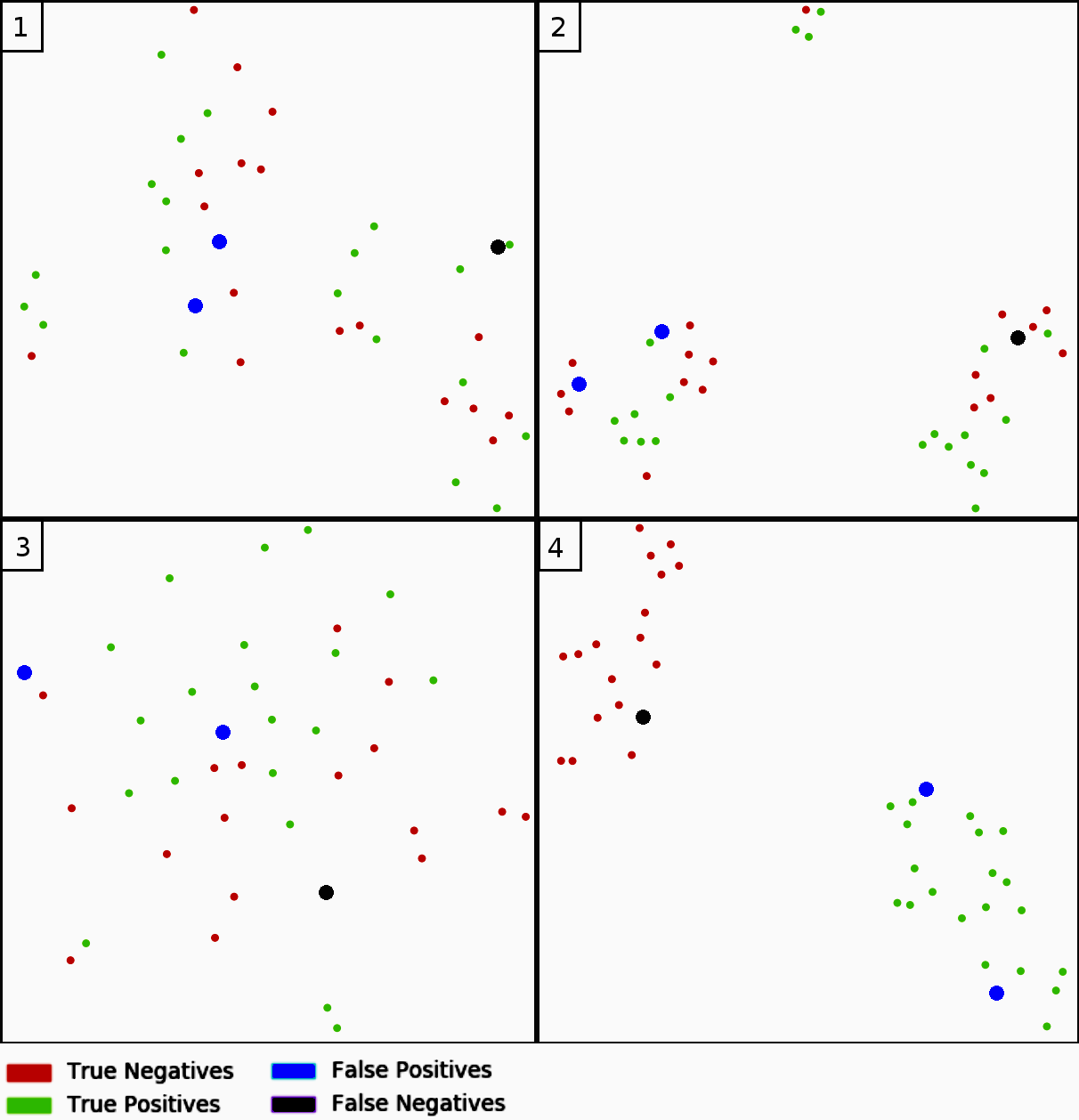}
\caption{\label{tsne}{\small{t-SNE embedding of the feature maps of 40
  validation images, at the input stage and at three intermediate
  stages of our network. 1) Concatenation of the pairs of input
  images, 2) Concatenation of the feature maps pairs at the end of the
  left and right branch, 3) Feature maps at the end of the Subtract
  layer, 4) Feature maps at the end of the second Dense
  layer. Negatives are "Stable" subjects, and Positives are "Decline"
  subjects. False positives and False negatives are represented by bigger circles for a better visibility.}}}

\end{figure}

Inspired by the work of Hosseini-Asl
et. al. \cite{hosseini2016alzheimer}, and to study in more depth the
discriminative abilities of our model, we plotted the t-distributed
Stochastic Neighbor Embedding (t-SNE) projection
\cite{maaten2008visualizing}, a nonlinear dimensionality reduction
algorithm, of the feature maps of 40 validation images obtained at
three stages of our network (see layers circled in red in
Fig.\ref{tsne}). The feature maps we chose are the concatenated
outputs of both left and right branches, the output of the Subtract
layer, and the output of the second Dense layer (see
Fig.\ref{graph}). We also plotted the concatenation of the input image
pairs for comparison. In these projections, similar feature maps are
drawn closer, and dissimilar feature maps are farther apart.

Fig.\ref{tsne} resumes the computing steps from the input step, square
number $1$ to the final result, square number $4$. We can observe at
the beginning, the formation of three clusters, square number $2$,
that are not probably related to our two classes "Stable" (Negative), and
"Decline" (Positive). We can 
hypothesize that they are due to other important characteristics of
the MRI images, that were extracted by the network, as the brain size
or another anatomical characteristics. Given that Bhagwat
et. al. used a concatenation layer to fuse the two branches, there
could be a bias introduced to their network at this step.

After the next layer, these three clusters are not identifiable
anymore (see Fig.\ref{tsne}.3), which could mean that the Subtract
layer had the effect of removing the unwanted features of our pair of
inputs.

Finally, at the penultimate Dense layer, the feature maps are grouped
in two clusters, corresponding to our two classes (see
Fig.\ref{tsne}.3). We showed false positives and false negatives in
additional colors to show how this embedding explains the
classification error of the model. Comparing this result to the plot
of the input pair, we can clearly see the discriminative ability of
our network.

\section{Conclusion}
In this paper we have proposed  3D-SiameseNet, a deep learning model
to analyze 3D brain MRI images in order to compare evolution between
two time points. The main contribution of this model is the use of raw
3D MRI data as input, suppressing the need for beforehand segmentation
and computation of manually selected features. Our architecture was
designed taking into account memory constraints as well as the need to
select local and precise features. By repeating the training and
validation phase 10 times, we obtain a mean validation accuracy of 90\%
after training.

In future works we would like to add clinical information to our
network to enhance its performance. The sub-network handling clinical
data will have to be designed to extract features from the raw
measures, before fusion with the imaging features.

\section*{Acknowledgements}
This project received the financial support from ICE project (RCB 2018-A02699-46) \\

Data   used   in   preparation   of   this   article   were   obtained   from   the  Alzheimer’s   Disease
Neuroimaging Initiative (ADNI) database (adni.loni.usc.edu). As such, the investigators
within the ADNI contributed to the design and implementation of ADNI and/or provided
data but did not participate in analysis or writing of this report. A complete listing of
ADNI       investigators       can       be       found       at:
http://adni.loni.usc.edu/wp-
content/uploads/how\_to\_apply/ADNI\_Acknowledgement\_List.pdf \\

Data  used  in  the  preparation  of  this  article  were  obtained
from  the  Alzheimer’s  Disease Neuroimaging  Initiative  (ADNI)
database. The primary goal of ADNI has been to test whether serial
magnetic resonance imaging (MRI),  positron  emission  tomography
(PET),  other  biological  markers,  and  clinical  and
neuropsychological  assessment  can  be  combined  to  measure  the
progression  of  mild cognitive impairment (MCI) and early Alzheimer’s
disease (AD).For up-to-date information, see www.adni-info.org. \\

Data   collection   and   sharing   for   this   project   was   funded   by   the  Alzheimer's   Disease
Neuroimaging  Initiative  (ADNI) (National  Institutes  of  Health  Grant  U01 AG024904)
and DOD ADNI (Department of Defense award number W81XWH-12-2-0012). ADNI is
funded by the National Institute on Aging, the National Institute of Biomedical Imaging
and   Bioengineering,   and   through   generous   contributions   from   the   following:  AbbVie,
Alzheimer’s  Association;   Alzheimer’s   Drug   Discovery   Foundation;   Araclon   Biotech;
BioClinica, Inc.; Biogen; Bristol-Myers Squibb Company; CereSpir, Inc.; Cogstate; Eisai
Inc.; Elan Pharmaceuticals, Inc.; Eli Lilly and Company; EuroImmun; F. Hoffmann-La
Roche Ltd and its affiliated company Genentech, Inc.; Fujirebio; GE Healthcare; IXICO
Ltd.;   Janssen  Alzheimer   Immunotherapy  Research   \&   Development,   LLC.;   Johnson   \&
Johnson Pharmaceutical Research \& Development LLC.; Lumosity; Lundbeck; Merck \&
Co., Inc.; Meso Scale Diagnostics, LLC.; NeuroRx Research; Neurotrack Technologies;
Novartis   Pharmaceuticals   Corporation;   Pfizer   Inc.;   Piramal   Imaging;   Servier;   Takeda
Pharmaceutical Company; and Transition Therapeutics. The Canadian Institutes of Health
Research   is   providing   funds   to   support  ADNI   clinical   sites   in   Canada.   Private   sector
contributions   are   facilitated   by   the   Foundation   for   the   National   Institutes   of   Health
(www.fnih.org).   The   grantee   organization   is   the   Northern   California   Institute   for
Research and Education, and the study is coordinated by the Alzheimer’s Therapeutic
Research Institute at the University of Southern California. ADNI data are disseminated
by the Laboratory for Neuro Imaging at the University of Southern California.

\bibliography{IEEEabrv,example}

\end{document}

%% file: preamble.tex
\usepackage[a4paper,textwidth=18cm,textheight=24cm,top=2.85cm, bottom=2.85cm, left=1.5cm, right=1.5cm]{geometry}

\usepackage{icdp2009}

\makeatletter
\long\def\@makecaption#1#2{%
\vskip\abovecaptionskip
\sbox\@tempboxa{#1. #2}%
\ifdim \wd\@tempboxa >\hsize
#1. #2\par
\else
\global \@minipagefalse
\hb@xt@\hsize{\box\@tempboxa\hfil}%
\fi
\vskip\belowcaptionskip}
\makeatother